\documentclass[journal]{IEEEtran}
\usepackage[style=ieee,maxnames=4,minnames=3,maxbibnames=3]{biblatex}

\usepackage{amsfonts}
\usepackage[cmex10]{amsmath}
\usepackage{amsthm}

\theoremstyle{definition}

\theoremstyle{remark}

\theoremstyle{remark}

\usepackage{booktabs}
\usepackage{array}
\usepackage{mdwmath}
\usepackage{mdwtab}
\usepackage{eqparbox}
\usepackage[caption=1]{caption}
\usepackage{stfloats}
\usepackage{url}
\usepackage{amssymb}
\usepackage{float}
\usepackage{indentfirst}
\usepackage{makeidx}
\usepackage{tabularx}
\usepackage{cases}
\usepackage{algorithmic}
\usepackage{mathtools}
\usepackage{color}
\usepackage{bm}
\usepackage{stfloats}
\usepackage{lipsum}
\usepackage{amsfonts}
\usepackage{bm}
\usepackage{dsfont}
\usepackage{lipsum}
\usepackage{graphicx}
\usepackage{url}
\usepackage{bbm}
 \usepackage{subcaption}
\newcommand{\RNum}[1]{\uppercase\expandafter{\romannumeral #1\relax}}
\ifCLASSOPTIONcompsoc
\usepackage[caption=false, font=normalsize, labelfont=sf, textfont=sf]{subfig}
\else
\usepackage[caption=false, font=footnotesize]{subfig}
\fi
\usepackage{multirow}
\allowdisplaybreaks
\makeatletter

\usepackage[linesnumbered,ruled,vlined]{algorithm2e}
\usepackage{amsmath}
\makeatother

\usepackage{amsthm}
\theoremstyle{definition} 
\theoremstyle{remark} 
\begin{document}
\title{Diffusion Models for Wireless Transceivers: From Pilot-Efficient Channel Estimation to AI-Native 6G Receivers}
\author{
    Yuzhi~Yang,~
    Sen~Yan,~
    Weijie~Zhou,~
    Brahim~Mefgouda,~
    Ridong~Li,~
    Zhaoyang~Zhang,~
    and~M\'erouane~Debbah
\thanks{Y. Yang, S. Yan, B. Mefgouda, and M. Debbah are with College of Computing and Mathematical Sciences, Khalifa University, Abu Dhabi 127788, UAE (e-mails: \{yuzhi.yang, sen.yan, brahim.mefgouda, merouane.debbah\}@ku.ac.ae). 
}
\thanks{W. Zhou, R. Li and Z. Zhang are with 1) College of Information Science and Electronic Engineering, Zhejiang University, Hangzhou 310027, China, and 2) Zhejiang Provincial Key Laboratory of Info. Proc., Commun. \& Netw. (IPCAN), Hangzhou 310027, China (e-mails: \{wj\_zhou, 12331116, zhzy\}@zju.edu.cn). 
}
}

\maketitle
\begin{abstract}
With the development of artificial intelligence (AI) techniques, implementing AI-based techniques to improve wireless transceivers becomes an emerging research topic. Within this context, AI-based channel characterization and estimation become the focus since these methods have not been solved by traditional methods very well and have become the bottleneck of transceiver efficiency in large-scale orthogonal frequency division multiplexing (OFDM) systems.
Specifically, by formulating channel estimation as a generative AI problem, generative AI methods such as diffusion models (DMs) can efficiently deal with rough initial estimations and have great potential to cooperate with traditional signal processing methods.
This paper focuses on the transceiver design of OFDM systems based on DMs, provides an illustration of the potential of DMs in wireless transceivers, and points out the related research directions brought by DMs. We also provide a proof-of-concept case study of further adapting DMs for better wireless receiver performance.
\end{abstract}

\section{Introduction}
In a typical wireless orthogonal frequency division multiplexing (OFDM) system, we typically model the transceiver as a signal recovery problem given the received signals. To solve the problem, how to efficiently utilize the prior knowledge of the transmitted signal and the channel becomes the main problem. Traditionally, we first obtain an initial channel estimation for the subcarriers with pilots, as the pilots are exactly known by the receiver. With some interpolation methods based on the channel modeling \cite{coleri2002channel}, we can further map the estimated partial channel matrices to the whole channel matrices. Further, the estimated channel is used to recover the transmitted signal with the help of the symbol prior distributions. We note that pilots and transmitted symbols share the same subcarriers in a resource block, and some semi-blind techniques can be used to combine all the signal priors together, which do not change the diagram above and are not the focus of this paper.

Revisiting the receiver diagram above in the artificial intelligence (AI) era, we have the following basic perceptions. First, how to utilize the signal prior is almost a solved problem. The modeling of signal distribution is mathematically strictly correct due to the existence of source coding. Taking such modeling as the precondition, existing Bayesian inference methods are optimal or suboptimal.
On the other hand, the channel matrices are highly structural while hard to be accurately described by explicit models \cite{11017513}. Existing works have also shown that AI methods usually remarkably outperform traditional ones on channel characterization problems \cite{csinet, tnse}.
However, the inference capability is still a critical challenge of neural networks (NNs). Also, the discreteness of the target symbols leads to another challenge of NN applications. Although the quantified NNs have shown good performance, their targets are usually not degrading too much by introducing quantification, while we need to utilize such sparsity in wireless systems.

Therefore, we are calling for an incorporated method of both kinds of methods.
Considering the combination of traditional and AI methods, there are existing works showing challenges in the adaptation of the interface \cite{10942479}. Most traditional signal processing methods are based on Bayesian inference, which requires a precise prior distribution instead of just an estimation as in most NN outputs. Another key challenge is that, in wireless transceivers, we can never obtain a noiseless estimation even for a small part of the channel matrices. Under different channel conditions, the transceivers should be adaptive to various pilot schemes and estimation SNRs, which are critical for non-generative NNs.

As a mainstream generative AI method, diffusion model (DM) offers more stable training and easier optimization than generative adversarial networks (GANs) or variational autoencoders (VAEs) \cite{ddpm}.
With the introduction of DMs, we can generate high-quality channel estimations, while traditional inference algorithms are retained for initialization and guidance \cite{MIMOdiffusion, 10930691, tnse, non-identical}. With the diffusion scheme, the NN can deal with any input pattern and input noise through appropriate embedding and guidance. Moreover, with the controllable error propagation in the diffusion scheme, it can cooperate well with traditional signal processing methods based on Bayesian inference. As introduced in \cite{tnse}, the DM-based receiver leads to a great performance gain and high adaptability compared to other AI receiver solutions. However, how to efficiently embed sparse prior estimation to guide the generation process still remains an important problem.

Specifically, we show the similarity of the channel recovery problem with signal inpainting and super-resolution reconstruction problems \cite{lugmayr2022repaint,yan2025diffnmr,yan2025diffnmr3,ldm}. In these well-investigated problems, DMs are employed to generate a complete, high-quality signal based on partial, low-quality observations. In the channel recovery problem, although the NN is working in a different domain, the task is still similar: generating full channel matrices from partial, noisy primary channel estimation results. Therefore, such methods are competitive candidates for the DM-based receiver.

In this paper, we discuss formulating the wireless transceiver through DMs. In particular, we aim to develop a novel wireless MIMO-OFDM transceiver based on a DM.
Specifically, we first discuss the necessity, overall framework, and basic requirements of the DM-based wireless transceiver. We then introduce several existing useful techniques on DMs, which have great potential to be used in the DM-based transceiver task. We further enumerate the important open problems in DM-based transceivers under the current protocols and new protocols, respectively. Moreover, we also conduct a case study of building a DM-based transceiver.

\section{Purposes of using DMs}
In this section, we first give several key reasons why adopting DMs instead of other AI methods in wireless receivers.
Firstly, we note that the transmitted symbols are generated after source coding. Thus, they do not have any structural properties other than the sparsity determined by the constellation diagram. Since we transmit irrelevant symbols over each subcarrier, the structural property of the received signals is not as strong as the channel matrices. Moreover, the commonly used transmission function and signal prior assumptions are accurate enough, and it is hard to improve them through NNs. Therefore, once we have obtained a channel estimation, the following procedure conducted through traditional algorithms is almost optimal. As NNs always provide suboptimal solutions, we should avoid applying them on problems we can already solve.

Therefore, in this paper, we only discuss applying NNs directly on channel characterization, which has not been completely solved by traditional methods yet.
The most important reason for using DMs instead of other AI frameworks is that DMs can deal with different input noise levels. In wireless communications, it is common that the channel quality varies over time in the same scenario. Thus, ensuring that the model can deal with different input qualities is essential, which DMs are good at.

Also, in DMs, the expected error power of the diffusion variable during the diffusion process is represented by the time indicator. Such an expected error is essential in most traditional wireless transceiver algorithms based on Bayesian inference, which is hard to obtain in other AI techniques.

Moreover, DMs can generate outputs that satisfy specific given input conditions—such as text, category labels, images, or even domain knowledge. We call it conditional guidance.
 In a wireless communication system, there are different observations from other transmissions that can assist the current one, such as the environment information, current user equipment (UE) location, and channel estimations from other time slots, frequencies, or locations. Such information is usually not sufficient for directly predicting the channel. However, it can be used to assist the estimation and reduce the necessary pilot overhead. To further exploit these indirect observations, a generative framework with guidance is required, which can be fulfilled by DMs.
\section{Preliminary DM Techniques}

\subsection{Preliminary of DMs}
\begin{figure}[t!]
    \centering 

    \begin{subfigure}[t]{0.45\textwidth}
        \centering
        
        \includegraphics[width=\linewidth]{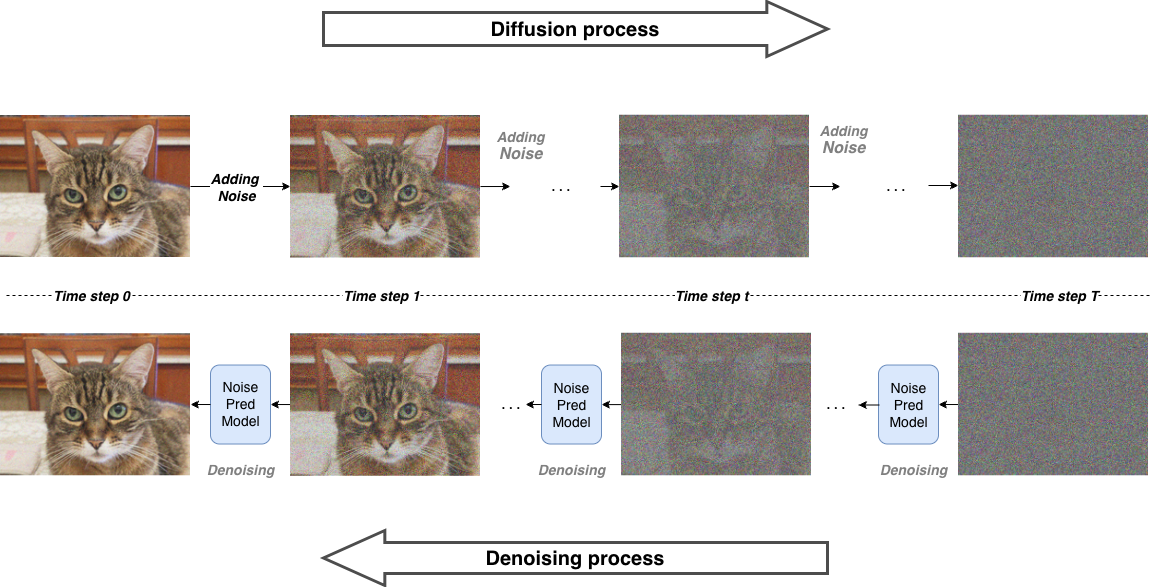}
        \caption{Diffusion Model Pipeline. Forward: the noise is gradually added to the input. Backward: starting from pure noise, the noise-prediction model progressively removes the noise.}
        \label{dm_pipe}
        
        \vspace{1em} 
        
        \includegraphics[width=\linewidth]{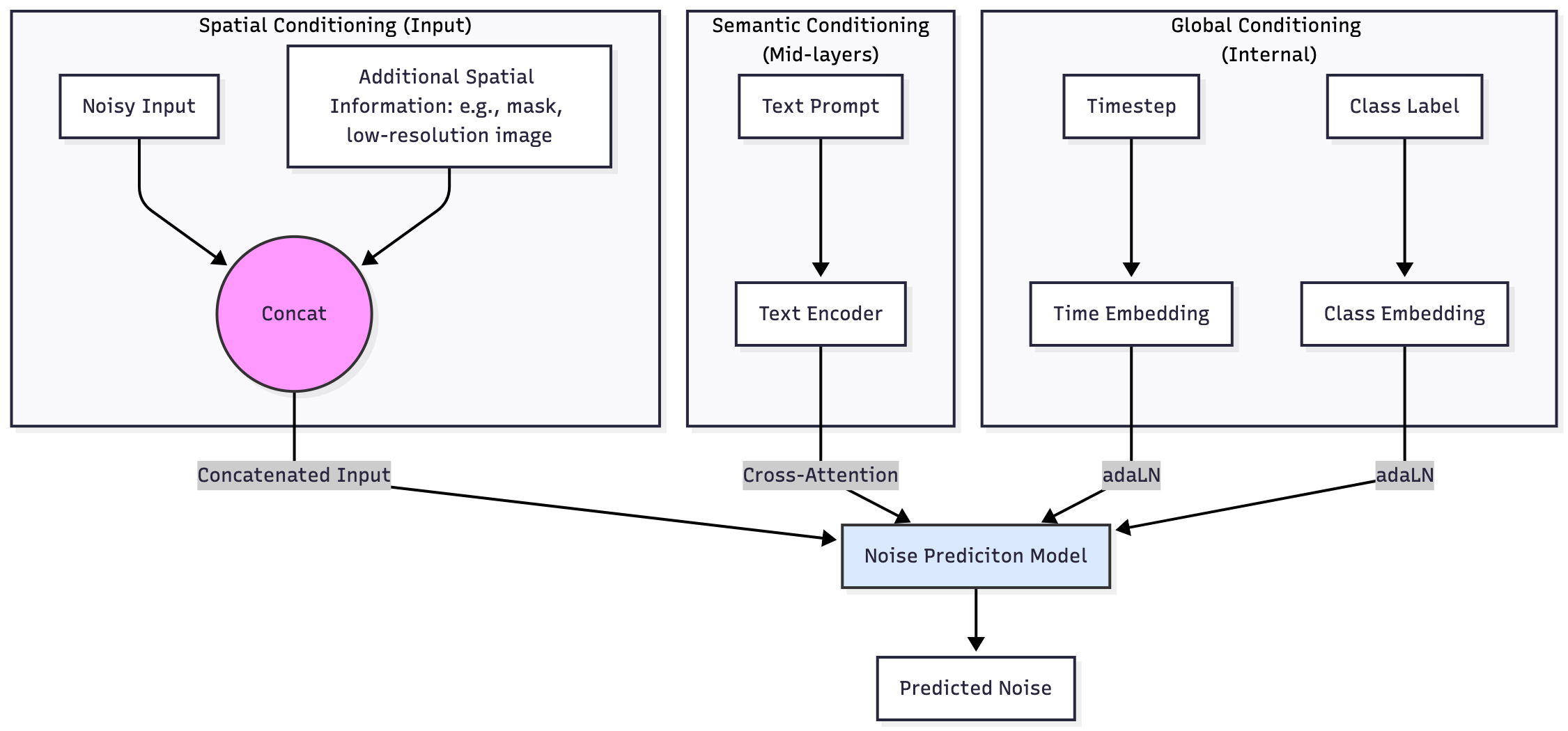}
        \caption{Representative Condition Mechanisms.}
        \label{unet_abstract}
        
    \end{subfigure}
    \vspace{1em}
    
    \begin{subfigure}[t]{0.45\textwidth}
        \centering
        \includegraphics[width=\linewidth]{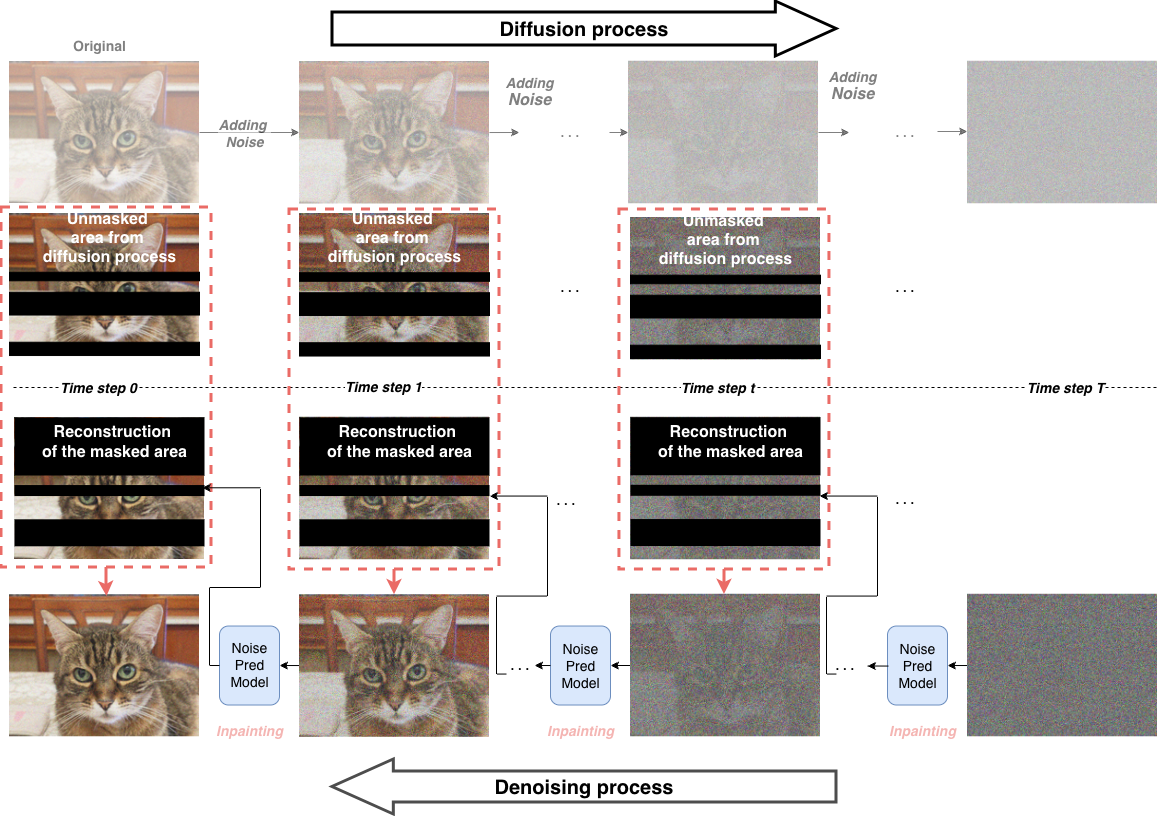}
        \caption{Repaint pipeline. It iteratively denoises the entire image while repeatedly resampling the known, unmasked areas during the reverse process to ensure consistency between the original content and the newly generated content.}
        \label{fig:repaint}
    \end{subfigure}
    \caption{Introduction of Diffusion model.}
    \label{fig:dm}
\end{figure}

DMs \cite{ddpm} have emerged as a powerful class of generative models for the downstream tasks such as data generation/synthesis, image editing, inpainting \cite{yan2025diffnmr} and super resolution \cite{ldm,yan2025diffnmr3}.

As illustrated in Fig. \ref{dm_pipe}, the core of these models lies in two key processes: the diffusion process (or called forward process) and the denoising process (or called reverse process). 
In the context of wireless signal recovery, the diffusion process is operated virtually by iteratively adding noise to the channel, and the model learns to inverse it. Like generating images through denoising, wireless channel estimation can also be formulated as denoising from a rough initial estimation.
In the actual receiver, we only operate the denoising process to recover the clean channel for further signal processing from initial estimation results and other auxiliary information.

\subsubsection{Diffusion Process}

The diffusion process gradually corrupts a clean signal by adding small amounts of noise in multiple steps. Starting from the original transmitted signal, the model progressively perturbs it over a series of steps until it becomes a completely noisy version. This process can be seen as a sequence of incremental randomizations, which helps the model learn how the signal evolves under noise and uncertainty. In wireless communication, this allows us to explicitly model the stochastic nature of channels and interference.

\subsubsection{Denoising Process}

The denoising process reverses the diffusion process by iteratively recovering the original data from noise. At each step, a NN (e.g., U-Net or transformer) is trained to predict and remove a fraction of the noise, progressively refining the signal. This approach simplifies generation by breaking down the complex task of direct sampling into a sequence of manageable denoising steps, offering a stable generation process. 

\subsubsection{Conditioning Mechanism} 
\begin{table*}[h!]
\centering
\caption{Conditioning Mechanisms in Representative Visual Downstream Tasks}
\label{tab:conditioning_mechanisms}
\setlength{\tabcolsep}{4pt}
\renewcommand{\arraystretch}{1.1} 
\begin{tabular}{@{} l p{3.8cm} p{7.8cm} @{}} 
\toprule
\textbf{Downstream Task} & \textbf{Typical Condition Input} & \textbf{Applicable Conditioning Mechanisms} \\
\midrule
Image Inpainting \& Super-Resolution &
\begin{itemize}
  \item Masked Image 
  \item Low-Resolution Image
\end{itemize} &
\begin{itemize}
    \item \textbf{Concatenation:} Conditional image is concatenated with the noisy input, providing direct spatial guidance. 
\end{itemize} \\
\midrule
Text-to-Image Generation &
\begin{itemize}
  \item Text Prompt 
  \item Class Label
\end{itemize} &
\begin{itemize}
  \item \textbf{Cross-attention:} Aligns generated content with text prompts.
  \item \textbf{Gradient-based Guidance:} Enhances image-text fidelity and sample quality.
  \item \textbf{Adaptive Layer Normalization (adaLN):} Injects class label information.
\end{itemize} \\
\midrule
Image Editing \& Translation &
\begin{itemize}
  \item Text Instruction 
  \item Style Image
\end{itemize} &
\begin{itemize}
  \item \textbf{Cross-attention} for text-based instructions.
  \item \textbf{Concatenation} for spatially aligned inputs (e.g., style images, semantic maps).
\end{itemize} \\
\bottomrule
\end{tabular}
\end{table*}

A key factor behind the success of DMs is their high degree of controllability. To steer the generation process towards a desired output that adheres to specific attributes or content, various conditioning mechanisms have been developed. These mechanisms effectively integrate external guidance, such as class labels, text descriptions, or reference images, into the iterative denoising process. As shown in Fig.~\ref{unet_abstract}, the primary mechanisms can be categorized as follows.

\textbf{Concatenation.} This is a straightforward approach where the conditional information is directly combined with the input to the denoising network. Typically, the condition (e.g., a feature map from a low-resolution image or a semantic map) is concatenated with the noisy image tensor along the channel dimension. This method is particularly effective for conditions that have a spatial structure aligned with the desired output.

\textbf{Gradient-based Guidance.} This technique uses the gradient from an auxiliary model to guide the sampling process. A classifier, pre-trained to recognize content in noisy images, calculates the gradient of the target class log-likelihood with respect to the current noisy image. This gradient, which points in the direction that makes the image more recognizable as the target class, is then used to perturb the reverse process update step. A popular variant, Classifier-Free Guidance, achieves a similar effect without needing an external classifier by training a single model on both conditional and unconditional inputs and then extrapolating the difference between their predictions.

\textbf{Cross-attention.} This mechanism is essential for integrating non-structural, sequential conditions like text. Within the denoising network (often a U-Net with attention modules), cross-attention layers are inserted to allow the model to attend to the most relevant parts of the conditional input (e.g., specific words in a text prompt) at each step of the denoising process. This provides a fine-grained and highly effective way to align the generated image with complex semantic instructions.

\textbf{Adaptive Layer Normalization (adaLN).} This method modulates the feature maps within the denoising network. Instead of learning static scale and shift parameters for normalization layers, these parameters are dynamically predicted by a small network whose input is a fusion of the timestep embedding and the conditional embedding (e.g., a class label vector). This allows the conditional information to influence the entire generation process at a deep feature level.

The application of these mechanisms often depends on the nature of the downstream task and the type of conditional input available. Table \ref{tab:conditioning_mechanisms} summarizes the relationship between key visual tasks and the conditioning mechanisms they typically employ. 
The image inpainting task is most analogous to the OFDM channel estimation task discussed in this paper.

\subsection{DMs for Inpainting Tasks}
Image inpainting is the task of reconstructing missing or damaged regions within an image. DMs excel at this by leveraging the surrounding context to generate content that is both semantically consistent and visually realistic. The model is conditioned on the masked image, where the known pixels serve as a strong guide for the reverse diffusion process. During sampling, the model only generates content for the unknown (masked) areas while ensuring the known (unmasked) areas are preserved.

\textbf{RePaint pipeline} \cite{lugmayr2022repaint} introduces an enhanced denoising strategy specifically for inpainting that improves coherence between the known and newly generated regions. Its pipeline distinguishes itself by using resampling to better enforce the condition. As shown in Fig.~\ref{fig:repaint}, the process for each denoising step is as follows.

Generation Step: A standard reverse diffusion step is performed to predict a less noisy image from the current noisy image.

Conditioning Step: The known regions in the generated image are replaced with their corresponding pixels from the original image (after adding the appropriate level of noise for the next timestep). This step re-imposes the ground truth information.

Resampling Step: Noise is added back to this "corrected" image to take it back to the starting image. This forward-then-backward resampling allows the model to harmonize the boundary between the generated and known regions, smoothing out inconsistencies before the next generation step.

The Repaint pipeline not only fully preserves information about known regions but also creates a more natural boundary between known and unknown regions.




\section{DM-based Transceiver Framework}
In this section, we then formulate the DM-based wireless transceiver structure and discuss its necessary requirements.

\subsection{System Overview}
\begin{figure*}[t]
  \centering
  \includegraphics[width=0.8\linewidth]{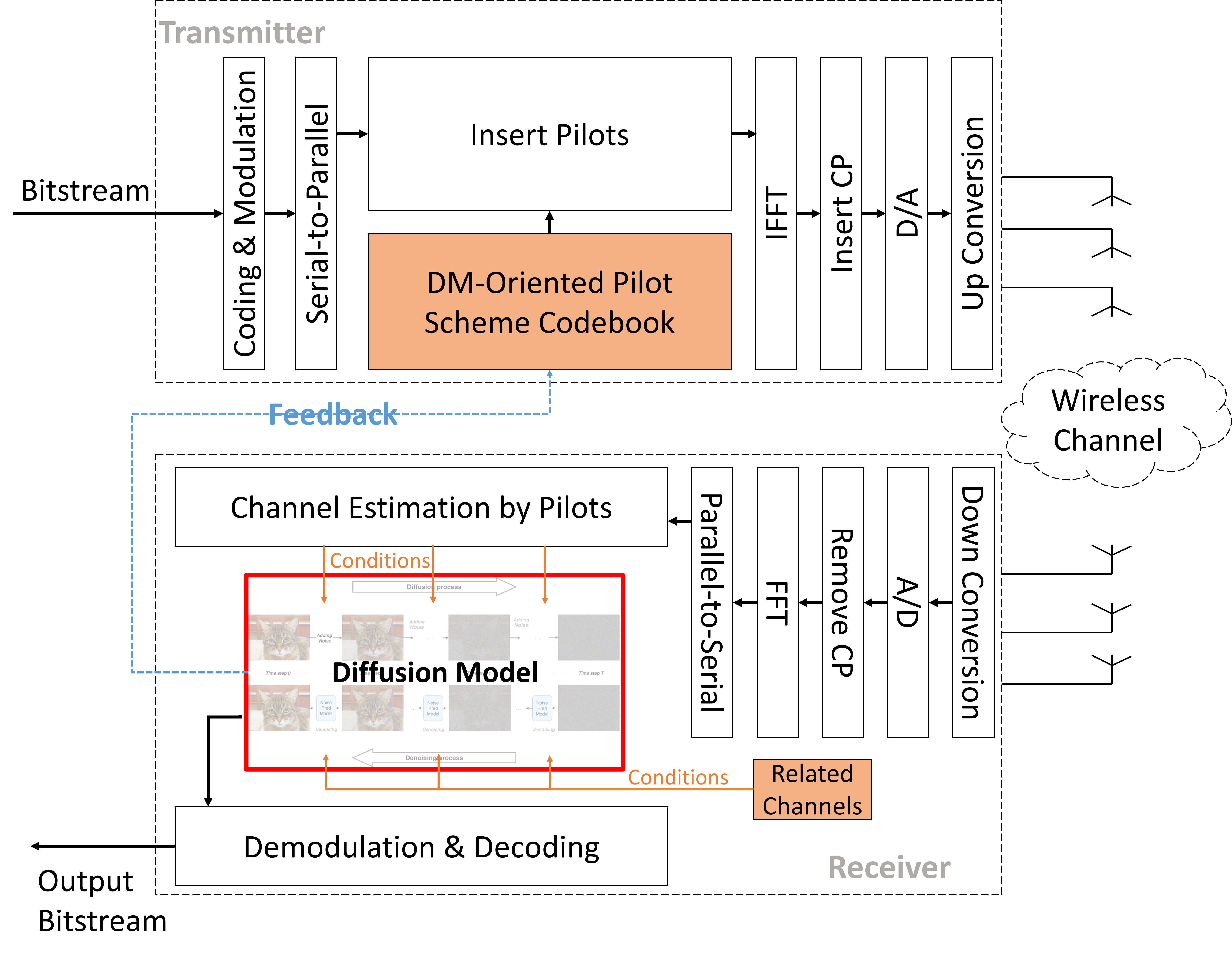}
  \caption{An overview of the transceiver design based on diffusion model.}
  \label{fig:overview}
\end{figure*}

With the context of DMs, we revisit the signal processing procedure of wireless receivers. Unlike the traditional pipeline of channel estimation, channel interpolation, and data estimation, we replace it with the iteration of condition-driven generation and verification, as illustrated in Fig. \ref{fig:overview}, where the corresponding transmitter is also illustrated.

Generally, most classical blocks are retained and become the backbone of the DM-empowered receiver, and similar key performance indicators (KPIs) are used. In the receiver, DM substitutes classical channel modeling and interpolation methods and produces channel candidates based on the original rough estimation. Alongside their classical role in the system, traditional demodulation and decoding methods based on Bayesian inference also critique the candidates and give feedback to the DM. Meanwhile, other related channel and scenario information also guides the DM through various conditioning mechanisms.

In the wireless receiver based on DMs, we begin with white noise or a very rough estimation. In each step, we employ the NN to enhance the generation quality. Under the basic hypothesis of DMs, we can always generate a channel matrix that follows the distribution of the dataset, which has the potential to substitute the interpolation modules.
However, unguided generation is likely to get a wrong channel, which may correspond to another position or scenario. Therefore, the information related to the signal prior should also be utilized to provide initialization, guide the diffusion process, and/or select better candidates to minimize hallucination.
\subsection{Technical Requirements}
Regarding the procedure described above, we summarize the key technical requirements of the DM specific to the wireless communication scenario. As indicated in table \ref{tab:2}, we show the main KPIs, supporting methods, and baseline methods for DM-based wireless receivers.
\begin{table*}[]
\caption{KPIs, supporting methods, and baseline methods for DM-based wireless receivers}
\label{tab:2}
\centering
\renewcommand{\arraystretch}{1.3}
\begin{tabular}{|l|l|l|}
\hline
                                 & Name                      & Description                                                                             \\ \hline
\multirow{3}{*}{Throughput KPIs} & Signal BER/BLER           & The ultimate goal of wireless receivers, while not directly related to the DM           \\ \cline{2-3}
                                 & Channel MSE/NMSE          & Direct criterion of DMs, strictly positively related to the BER/BLER                    \\ \cline{2-3}
                                 & Pilot Density             & Lower pilot density leads to better spectrum efficiency with the cost of channel MSE    \\ \hline
\multirow{3}{*}{Latency KPIs}    & Backbone NN Complexity    & Computationally efficient backbone NN is essential to conduct DM in reasonable time     \\ \cline{2-3}
                                 & Denoising Steps           & Using fewer denoising steps or efficient denoising pipeline sharply reduces the latency \\ \cline{2-3}
                                 & Caching / Parallelization & Caching / Parallelization methods can also reduce the inference time overhead           \\ \hline
\multirow{3}{*}{Algorithm Units} & LMMSE, AMP, etc.          & Traditional signal processing methods to be integrated to the DM-empowered receiver     \\ \cline{2-3}
                                 & Channel Interpolation NNs & Low-complexity channel characterization NNs are candidates of DM's backbone             \\ \cline{2-3}
                                 & Channel Mapping NNs       & Using channel mapping NNs to provide conditions by related channel matrices             \\ \hline
\multirow{2}{*}{Baselines}       & Channel Interpolation     & Simple linear interpolation / interpolation based on channel modeling / NN solutions    \\ \cline{2-3}
                                 & Other GenAI Methods       & Classical generative AI solutions based on VAE, GAN, etc.                               \\ \hline
\end{tabular}
\end{table*}
\subsubsection{Guidance Based on Partial Observations}
The first requirement is brought by the partial estimation. In a wireless resource block, there exist pilot and symbol parts. By the symbol prior of sparsity determined by the modulation scheme, we can use the symbol parts to improve the channel estimation through blind or semi-blind estimation methods. However, they are always of secondary importance compared to the pilot-based estimations, which are far more informative.

If we focus solely on the pilot scheme, the receiver's task becomes generating a full channel estimate from partial and potentially noisy pilot observations. This aligns conceptually with partial observation generation tasks explored in computer vision\cite{lugmayr2022repaint,yan2025diffnmr}.

\subsubsection{Insertion of Traditional Signal Estimation Algorithms}


Traditional signal processing methods are crucial in traditional transceivers. They often provide almost optimal solutions given precious channel estimates and are good at dealing with inference-based systems with explicit modeling and cannot deal with implicit scenarios well, which are exactly in contrast to NNs. 

Therefore, we need to incorporate NN with traditional signal estimation algorithms in the intelligent transceivers as wireless transmission is a complicated system with both implicit and explicit parts.
One straightforward method is to provide guidance from the traditional methods to the NN indirectly through feedback. However, a crucial problem exists that most existing signal estimation methods are non-differentiable while most NN-related feedback methods are based on gradient algorithms. Therefore, it becomes an important problem to design appropriate insertion of non-differentiable units into the diffusion framework.

\subsubsection{Adaptation to Different Transmission Schemes}

Unlike traditional methods where we can adaptively adjust the parameters, NN-based frameworks require training for each channel condition to adapt to different transmission schemes.


With the design in Fig. \ref{fig:overview} which separates the channel and signal estimation parts, the compatibility regarding symbol attributes including modulation and channel coding can be stripped from the NN-related parts and realized by traditional methods.
Thus, specific to the diffusion backbone, the most important compatibility requirements lie in different pilot density and noise levels. These factors directly impact the reliability of the initial pilot-based estimate. Also, we always seek for the minimum pilot density while maintaining the error rate in the wireless system designs. This implies that there is no universal pilot scheme applicable to all devices, whereas an adaptive pilot scheme is required.
Both issues are intrinsic and unavoidable requirements brought by wireless communications. Thus, to realize DM-based transceivers, both issues must be settled well within the same NN.

\subsubsection{Complexity Limitations}
Another basic requirement lies in the complexity. To meet the ultra-reliable low-latency communication (URLLC) requirement, it is unaffordable to employ high-complexity components in wireless transceivers, although they may perform better.
Especially for the DMs, more generation steps usually bring refined results, while the performance gain is not linear. Therefore, the tradeoff between the computational complexity and overall performance also requires attention.

\section{Open Problems on DM-based Wireless Transceivers}
\begin{figure*}[t]
  \centering
  \includegraphics[width=0.95\linewidth]{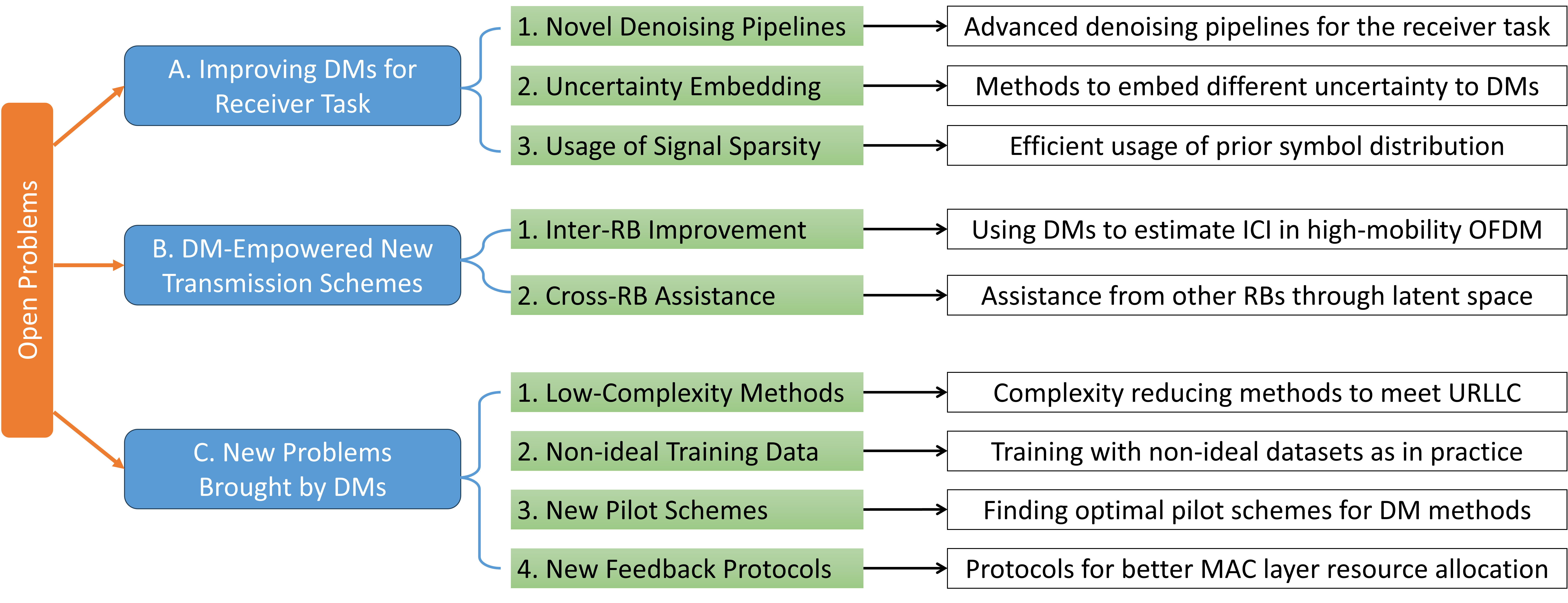}
  \caption{An overview of open problems of applying DMs in wireless transceivers.}
  \label{fig:open_probs}
\end{figure*}

In this section, we discuss the urgent research topics and open problems for DM-based receivers. An overview of the open problems discussed in this section is shown in Fig. \ref{fig:open_probs}.

\subsection{Improving DMs for Receiver Task}
Although some existing works \cite{MIMOdiffusion, tnse, non-identical} have built an illustration of how DM-based receivers look like, there still remain a few important topics for future works. In this section, we focus on the existing transmitting scheme of multiple input multiple output (MIMO)-OFDM. In particular, we enumerate several research directions on realizing the wireless receiver by DMs, where the DMs are only used to replace the channel estimation units in current systems.
\subsubsection{Novel Denoising Pipelines}


Apart from the vanilla diffusion, there are two common diffusion generation pipelines for image recovery problems as introduced above, namely the inpainting and super-resolution methods. However, when applied to the channel task, they still have remarkable problems to overcome. Wireless channels have strong structural features and sparse pilot observations. Crucially, channel estimation at the pilot positions is not completely reliable due to the noise, and they only indicate areas of higher relative reliability compared to unobserved parts, differing from the often clean observations in computer vision tasks. Thus standard inpainting, which fixes known regions during generation, is unsatisfactory because it ignores this inherent noise and unreliability in the channel estimation task. Similarly, treating the task as super-resolution suffers from weak guidance. Due to the sparsity of pilots and the noisy estimations, we cannot provide guidance as powerful as conventional tasks in a wireless receiver, which is also indicated in \cite{tnse}.

\subsubsection{Embedding Uncertain Observations}


While DMs typically use time embeddings to parse the uncertainty to the NN and binary masks for known/unknown regions in tasks like inpainting and super-resolution, these methods are insufficient for wireless tasks where different elements have varying uncertainties. Thus, a richer representation like soft masking is needed. Although our prior work \cite{non-identical} proposed a novel uncertainty embedding scheme for dimension-wise wireless resource allocation, it did not focus on guidance. We can naturally extend the binary mask concept to soft masks based on confidence levels from signal processing modules, allowing further investigation into mask embedding mechanisms in DMs, which may feedback DM research.

\subsubsection{Usage of Prior Data Distributions}

Beyond prior channel structure, wireless transceivers leverage symbol prior knowledge to enhance channel estimation. However, the commonly applied methods based on Bayesian inference are not compatible with NNs. Thus, how to effectively utilize such data priors becomes a crucial problem in DM-based transceivers. Actually, we can come up with the following three paths. The first is to adapting traditional inference algorithms such as expectation-maximization for imperfect NN inputs. Secondly, we can embed the symbol estimations to guide the NNs. Another method is to indirectly exploit priors via the criterion, as signal processing components can identify high-quality candidates during the generation as shown in the paper \cite{tnse}.

\subsection{DM-Empowered New Transmission Schemes}
In this section, we further discuss the transmission scheme, and focus on novel transmission schemes brought by DMs. One main benefit of DM-based structures is that we do not necessarily need high-quality initial estimation based on pilots. Traditionally, the channel estimation results obtained through pilots determine the performance upper bound of the following signal processing. However, in DM-based frameworks, we only need them to provide guidance, making it possible to use fewer pilots even insufficient for independent partial channel recovery.

\subsubsection{Inter-RB Improvement of OFDM Under High-Mobility Scenarios}

In the current OFDM scheme, we usually make the stable assumption. It has been proven that the OFDM scheme can solve all inter-carrier interference (ICI) and inter-symbol interference (ISI) issues with the introduction of cyclic prefix (CP) under the assumption that the scenario is invariant within the symbol. However, when considering the Doppler effect, which is widely existing in practice, there exists ICI even after frequency compensation due to the multi-path phenomenon.
Such ICI has become the bottleneck of current OFDM applications, especially in high-mobility scenarios.

Alongside the traditional methods such as OTFS, AI has provided us another method to deal with such ICI issues by embracing the Doppler effect. That is, we acknowledge the existence of the Doppler effect and try to model it with NNs. By considering the ICIs as a part of the OFDM channel, it no longer degrades the receiver performance.
However, such modeling introduces another dimension of the wireless channels, making it hard to conduct channel estimation. Typically, to estimate the ICI on each subcarrier, several adjacent pilots are required, introducing an unaffordable overall pilot overhead.

Meanwhile, DMs can generate the channel by initializing with rough estimation even without prior ICI knowledge.
Iterative denoising progressively refines the channel estimation while leveraging symbol sparsity priors through traditional signal processing to extract ICI information, which is then fed back to guide subsequent denoising iterations. Thus, with the introduction of DMs, it becomes possible to completely solve the ICI problem in the OFDM systems with minimum changes.

\subsubsection{More Abundant Cross-RB Guidance}
If we consider the transceiver from a high-level perspective other than focusing on a RB, we can find other observations related to the target resource block. As widely shown in the existing literature, the channel matrices corresponding to different frequencies, near locations in the same scenario, and other historical records having similar conditions can be used to predict the target channel matrix. Although these methods may perform poorly when applied alone, there is great potential to integrate them with the pilot-based estimation to further reduce the necessary pilot costs.
As also illustrated in many provision papers on wireless large models \cite{zhu2025wirelesslargeaimodel}, such integration has the potential to be realized through a unified large model. In the existing large language models, we can align the language and images to the same latent domain, and use some techniques like stable diffusion to generate images according to the text inputs. Similarly, via wireless large models, we can use the observations and estimations from other transmissions to assist and lead the DM-based generative wireless receiver.

\subsection{New Problems Brought by DMs}
Alongside the introduction of DMs, there also arise new problems in the transceiver design, and some protocols require reconsideration, which are discussed in this section.
\subsubsection{Low-Complexity Network Topologies and Algorithms}


Most existing works assume much larger OFDM subcarrier spacings than those in practice. Although larger spacings increase the difficulty of the interpolation problem, practical wide-band systems use much smaller spacings, which introduce many more subcarriers. This sharply enlarges the problem scale, leading to high complexity in architectures such as cross-attention and requiring more training data. Additionally, considering computational constraints, applying knowledge distillation in DMs can reduce the required diffusion steps and overall computational cost, helping meet strict URLLC requirements.

\subsubsection{Training with Non-Ideal Data}

Another practical problem that is seldom discussed in the existing works lies in the data acquirement. Most existing works conduct experiments on simulated datasets. Since existing channel modeling methods such as ray-tracing have shown high accuracy in characterizing realistic scenarios, such datasets are enough to prove the effectiveness of such methods. However, as the model needs to learn about the scenario information, practical data after deployment is also essential for intelligent communication devices.
However, in practice, we can never obtain ideal channel matrices due to the noise. Instead, we usually only have partial and noisy channel matrices from the communication history. How to use these poor-quality data to train or finetune the DM remains to be another problem.

\subsubsection{Optimal Pilot Schemes for DM Transceivers}

The introduction of DMs also requires a revisit of the optimal pilot schemes.
Traditional pilot schemes include block, comb, and lattice-type pilots and we use equally spaced pilots for better channel interpolation, and use adjacent pilots to deal with the Doppler effect.
However, such straightforward principles do not necessarily hold for NNs, requiring further study on the optimal pilot schemes provided the pilot density.

\begin{figure*}[t]
\centering
  \begin{subfigure}[b]{0.9\linewidth}
    \centering
  \includegraphics[width=1\linewidth]{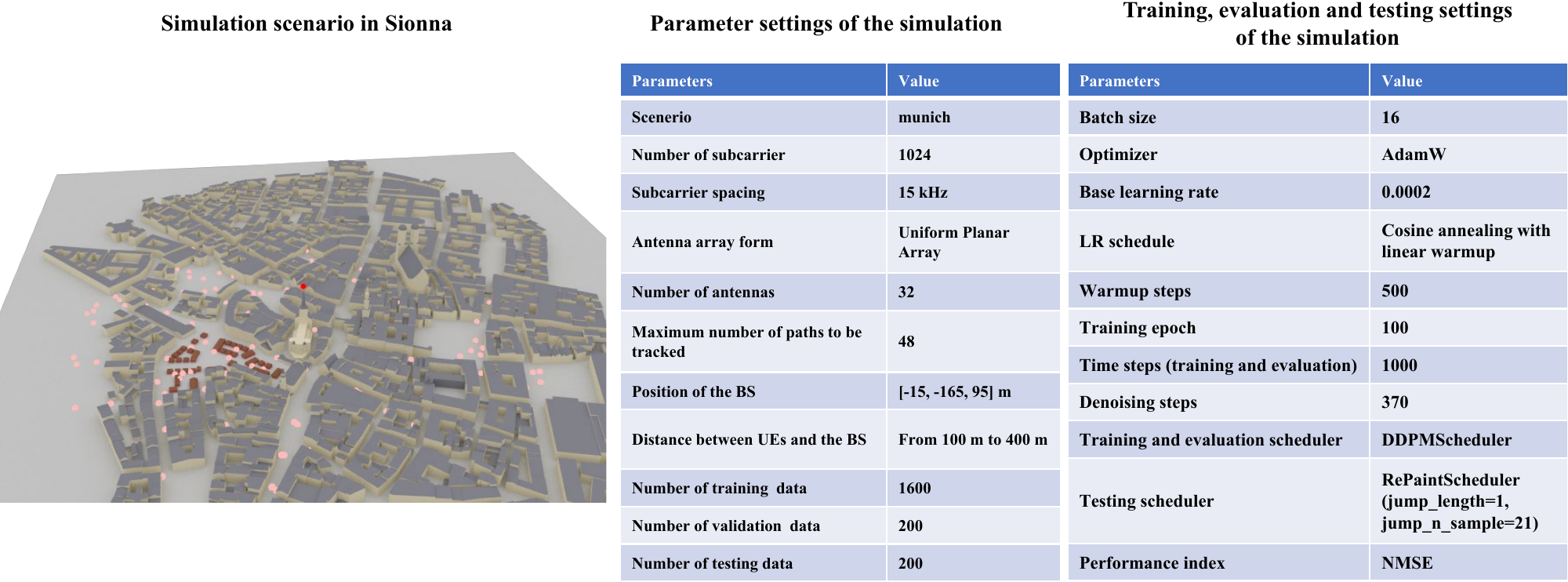}
  \caption{The scenario graph and the parameter settings of the experiments. The left figure shows the simulation scenario in Sionna, where the red point on the top of the tower represents the BS and the pink points represent the UEs. }
  \label{fig:scene_parameter}
  \end{subfigure}
  
  \begin{subfigure}[b]{0.9\linewidth}
    \centering
    \includegraphics[width=\textwidth]{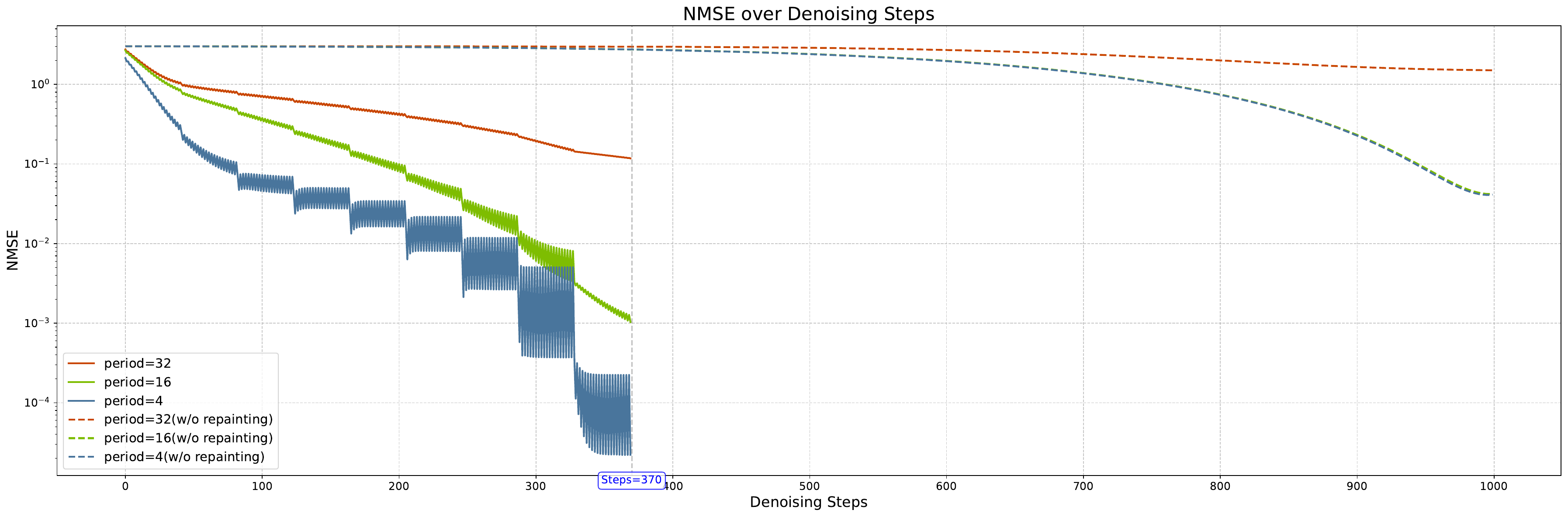}
    \caption{NMSE performance curves for different pilot density in the denoising process.}
    \label{fig:mse_nmse_plot}
  \end{subfigure}
  
\begin{subfigure}[b]{0.9\linewidth}
    \centering
    \includegraphics[width=\textwidth]{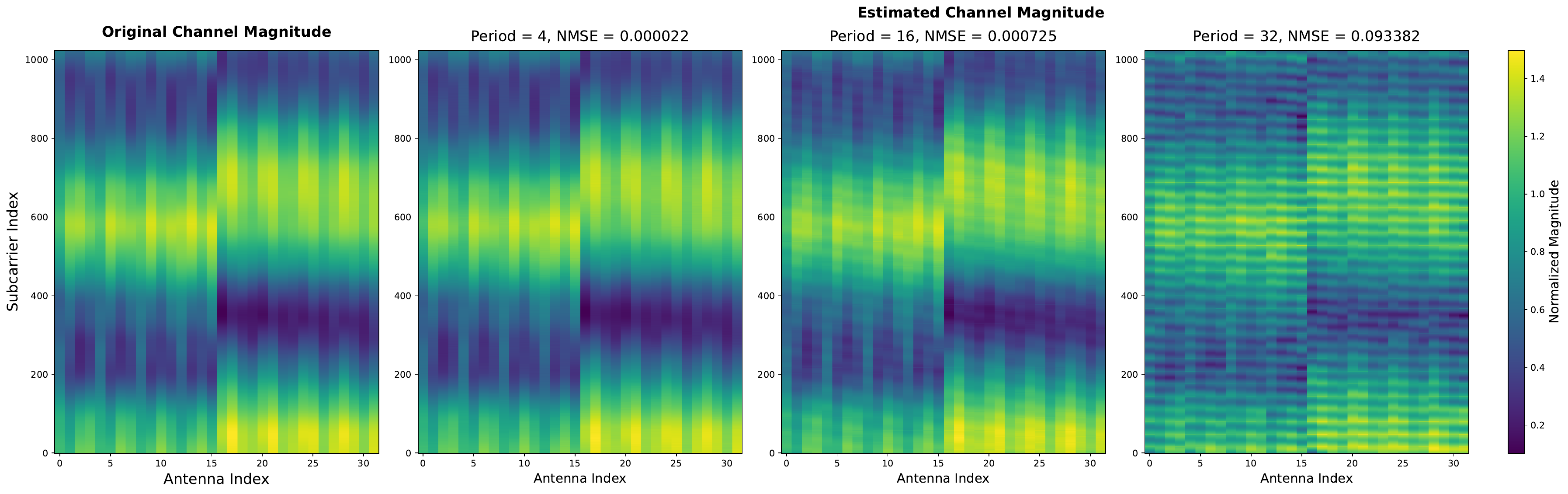}
    \caption{Comparison of channel estimation results for different pilot density with repaint pipeline.}
    \label{fig:fig_GT_est}
  \end{subfigure}
  
  \caption{Performance of channel estimation under different pilot density versus denoising steps with and without the repaint pipeline.}
  \label{fig:results}
\end{figure*}


Specifically, for DM-based methods, it remains unclear whether equally spaced pilots are still optimal. Since DMs extract channel characteristics directly from data rather than explicit models, pilots with nonuniform distributions might capture richer multi-scale features, requiring further investigation through massive simulations under different scenarios. As we can evaluate the channel estimation quality without the ground truth by the symbol prior distribution, the optimal pilot scheme
can even be chosen based on previous transmissions.

Moreover, in multi-user scenarios, the pilot scheme requires even further study. DMs can leverage pilots indirectly even without explicit initial estimates, enabling greater pilot reuse among UEs, which is an appealing direction for future research.

\subsubsection{Training and Feedback Protocols}

NN-based transceivers are less interpretable and require training before deployment, making new communication protocols necessary. In traditional algorithms, the impact of each parameter is clear and performance is controllable, but NN behavior is unpredictable before real testing. Moreover, the DM also introduces a few parameters such as the diffusion steps, which also require adjustment over different scenarios. Hence, intelligent transceivers need exploratory protocols for initial access and adaptive feedback during transmission.



Moreover, data collection, training, and fine-tuning should be integrated into new protocol designs. The related efficiency and privacy issues are worth further discussion.  Another issue lies on the user side. Given the diversity of wireless users, not all devices can or will employ NNs. Therefore, different protocols are required for UEs that cannot use NNs, can use pre-trained NNs, or are capable of on-device training.

\section{Case Study}

In this section, we take a demo experiment to preliminarily demonstrate the feasibility of the repaint pipeline for the channels estimation task based on the pilots. For our experiments, we use the Sionna \cite{sionna} as the channel simulator to generate training, testing, and validation data in the scene containing the area around the Frauenkirche in Munich, Germany based on the ray-tracing. Fig. \ref{fig:scene_parameter} details this demo experiment, including scenario graph, and parameter tables. UEs with a signal antenna are randomly distributed in the scenario, and their maximum distance from the BS corresponds to entry "Maximum distance between UEs and the BS" in the table. Each UE's Channel Frequency Response (CFR) with the BS serves as a sample. We compare the channel estimation results of the DM across different pilot density. In our experiments, we conduct training on datasets corresponding to each pilot density individually. Specifically, one pilot is inserted every 4, 16, or 32 subcarriers, with pilot positions randomly chosen within each interval yet uniformly spaced overall. We use a 0-1 mask to indicate the pilot positions, where 1 denotes that a pilot is inserted in this positions. 


Since CFR is complex-valued, we concatenate its real and imaginary parts along the channel dimension.  During the diffusion process, the channel estimation results at the pilot positions, the inverse mask, and the channel corrupted by complex Gaussian noise in a random time step are concatenated along the channel dimension as the input to the U-Net. During the denoising process, the channel estimation results at the pilot positions, the inverse mask, and the complex Gaussian noise are concatenated along the channel dimension as the initial input to the U-Net. Furthermore, we employ the repaint pipeline \cite{lugmayr2022repaint} to better condition the signals. 

Numerical results are shown in the curves graphs of Fig. \ref{fig:mse_nmse_plot} and the amplitude of the CFR generated under 370 denoising steps with the different pilot density are shown in Fig. \ref{fig:fig_GT_est}. Based on the results, even under low pilot density conditions, the DM model still yields low-NMSE channel estimation results and exhibits excellent performance in the inpainting task. Additionally, the repaint pipeline enables the model to achieve better performance and faster convergence with fewer denoising steps. Considering the inference time overhead, reducing the number of inference steps, while causing a slight performance drop, still maintains high estimation accuracy. This explains why we use fewer steps in the denoising process than in the diffusion process. Furthermore, this trade-off depends on the requirements for inference latency, pilot density, and estimation accuracy.

\section{Conclusion}
In this paper, we discussed the potential usage of DMs in wireless receivers and relevant problems brought by them in transceiver designs. We illustrated the overall workflow of DM-based transceivers and introduced some relevant DM techniques.
Further, we provided an overview of the new problems lying in DM-based receivers. Specifically, we discussed several potential adaptations to existing DM techniques for wireless transceivers, new transmission schemes brought by the introduction of DMs, and new protocol-level designs for the DM-based receivers. Our case study also showed the great potential of DM applications in wireless transceivers and verified the possibility of developing transceiver-specified DM techniques.

\printbibliography

\end{document}